\documentclass[11pt,twoside]{article}
\usepackage{asp2010}

\resetcounters

\begin{document}

\title{Modelling colliding wind binaries with RAMSES, extension to special relativity}

\author{Astrid Lamberts,$^1$  S\'{e}bastien Fromang,$^2$ Guillaume Dubus,$^1$ and  Geoffroy Lesur$^1$
\affil{$^1$ UJF-Grenoble 1 / CNRS-INSU, Institut de Plan\'{e}tologie et d\textquoteright Astrophysique de Grenoble (IPAG)\\ UMR 5274, Grenoble, F-38041}
\affil{$^2$ Laboratoire AIM \\CEA/DSM-CNRS-Universit\'{e} Paris Diderot, IRFU/Service d\textquoteright Astrophysique \\CEA-Saclay F-91191 Gif-sur-Yvette, France}}

\begin{abstract}
We present simulations of colliding supersonic stellar winds with RAMSES. The collision results in a double shock structure that is subject to different instabilities.Properly modelling these instabilities requires a high enough resolution. At large scale, orbital motion is expected to turn the shocked zone into a spiral but we find that in some configurations the Kelvin-Helmholtz instability (KHI) may disrupt the spiral. A colliding wind structure is also expected in gamma-ray binaries composed of a massive star and a young pulsar that emits a highly relativistic wind. Numerical simulations are necessary to understand the geometry of such systems and should take into account the relativistic nature of the pulsar wind. We implemented a second order Godunov method to solve the equations of relativistic hydrodynamics (RHD) with RAMSES, including the possibility for Adaptive Mesh Refinement (AMR). After a brief overview of our numerical implementation, we will present preliminary simulations of gamma-ray binaries.

\end{abstract}

\section{Introduction}

Massive stars posses highly supersonic winds. In binary systems, the interaction of two winds creates a double shock structure which geometry depends on the momentum flux ratio of the winds. The winds are separated by a contact discontinuity. Numerical simulations show that several instabilities may arise in the colliding wind region \citep{2009MNRAS.396.1743P}. For isothermal winds, the non-linear thin-shell instability \citep{1994ApJ...428..186V} dominates the collision region while the KHI can strongly perturb the shocked region between adiabatic winds \citep{PaperI}.

Up to now, high-resolution simulations have focused on the region close to the binary. At larger scale orbital motion turns the shocked structure into a spiral. Matter is expected to be ballistic, with a distinction between both shocked spiral arms \citep{2011ApJ...726..105P}. The exact structure of the spiral is still to be studied. A spiral structure is observed in WR 104, a binary composed of a Wolf-Rayet (WR) and an early-type star. Owing to dust emission, it displays a spiral in infrared up to a few hundred times the binary separation \citep{2008ApJ...675..698T}. The formation of dust in such systems is still poorly understood, and is likely to be facilitated in the colliding wind region. We perform simulations of this system to determine the structure of the colliding wind region and put constrains on dust formation.

$\gamma$-ray binaries are binaries composed of a massive star and a young pulsar (see e.g. \citet{2009ApJ...706L..56A}) with a tenuous, highly relativistic wind. The collision between the pulsar wind and the wind from the companion star, which results in $\gamma$-ray emission, is expected to present similarities with the collision between stellar winds \citep{2006A&A...456..801D}. Still, the relativistic nature of the pulsar wind is likely to modify the dynamics and emission properties of large scale colliding wind region that can be observed in radio (see e.g. \citet{2011ApJ...732L..10M}). To investigate the impact of the relativistic nature of the pulsar wind, we developped a relativistic extension to the hydrodynamical code RAMSES \citep{Teyssier2002}. We explain our numerical implementation, focusing on the AMR and present some preliminary simulations of $\gamma$-ray binaries.

\section{Large scale structure of colliding wind binaries}
We use the hydrodynamical code RAMSES for our simulations. It is a second order Godunov method, that solves the equations of hydrodynamics. We focus on the adiabatic limit. We use AMR to increase resolution at the shocks and properly model the instabilities while simulating the large scale structure at reasonable computational cost. The winds are generated following the method by \citet{Lemaster:2007sl}. We include a passive scalar to distinguish the winds and determine mixing.

\begin{figure}[h]
  \centering
  \includegraphics[width = .3\textwidth]{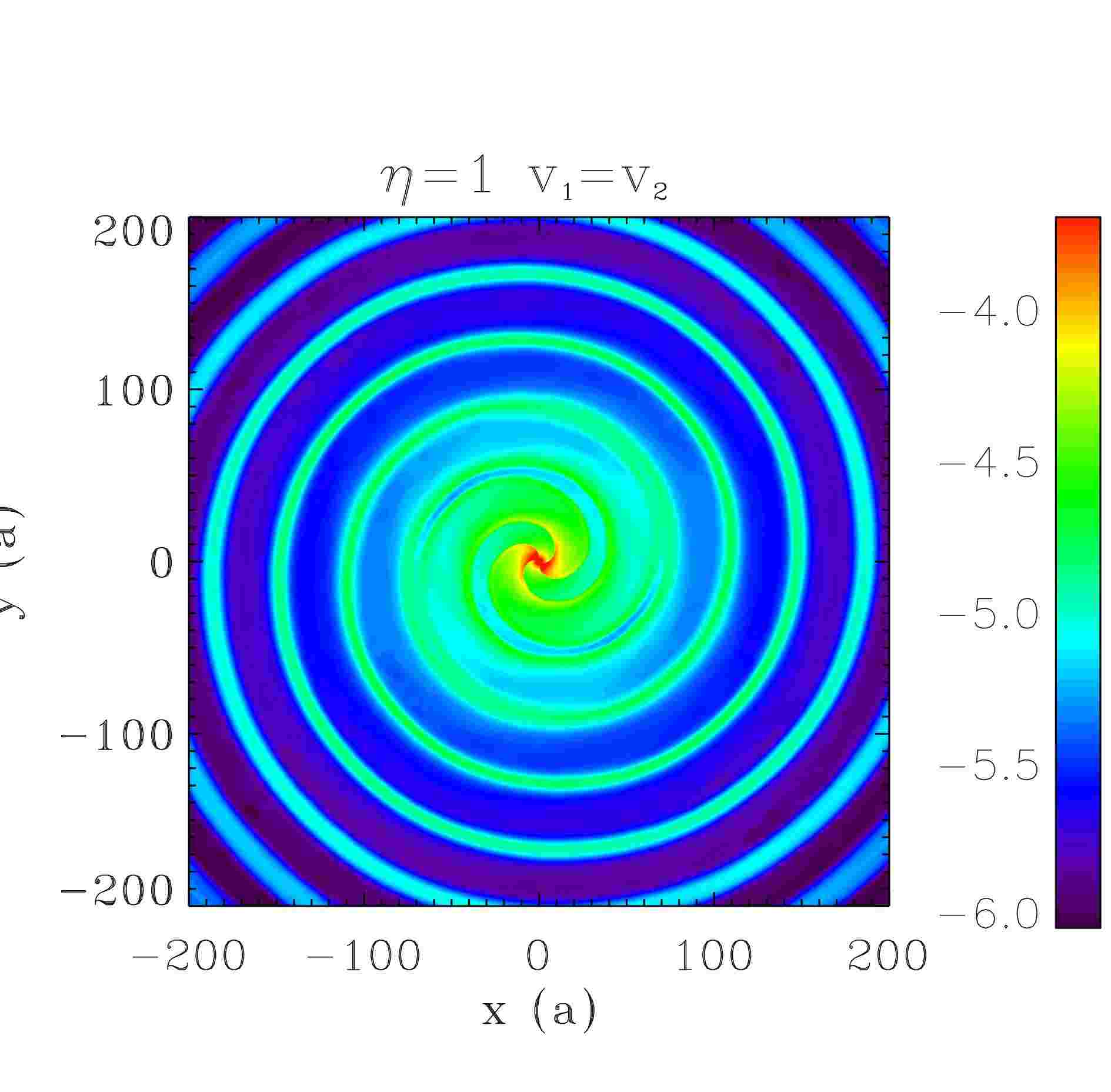}
  \includegraphics[width = .3\textwidth]{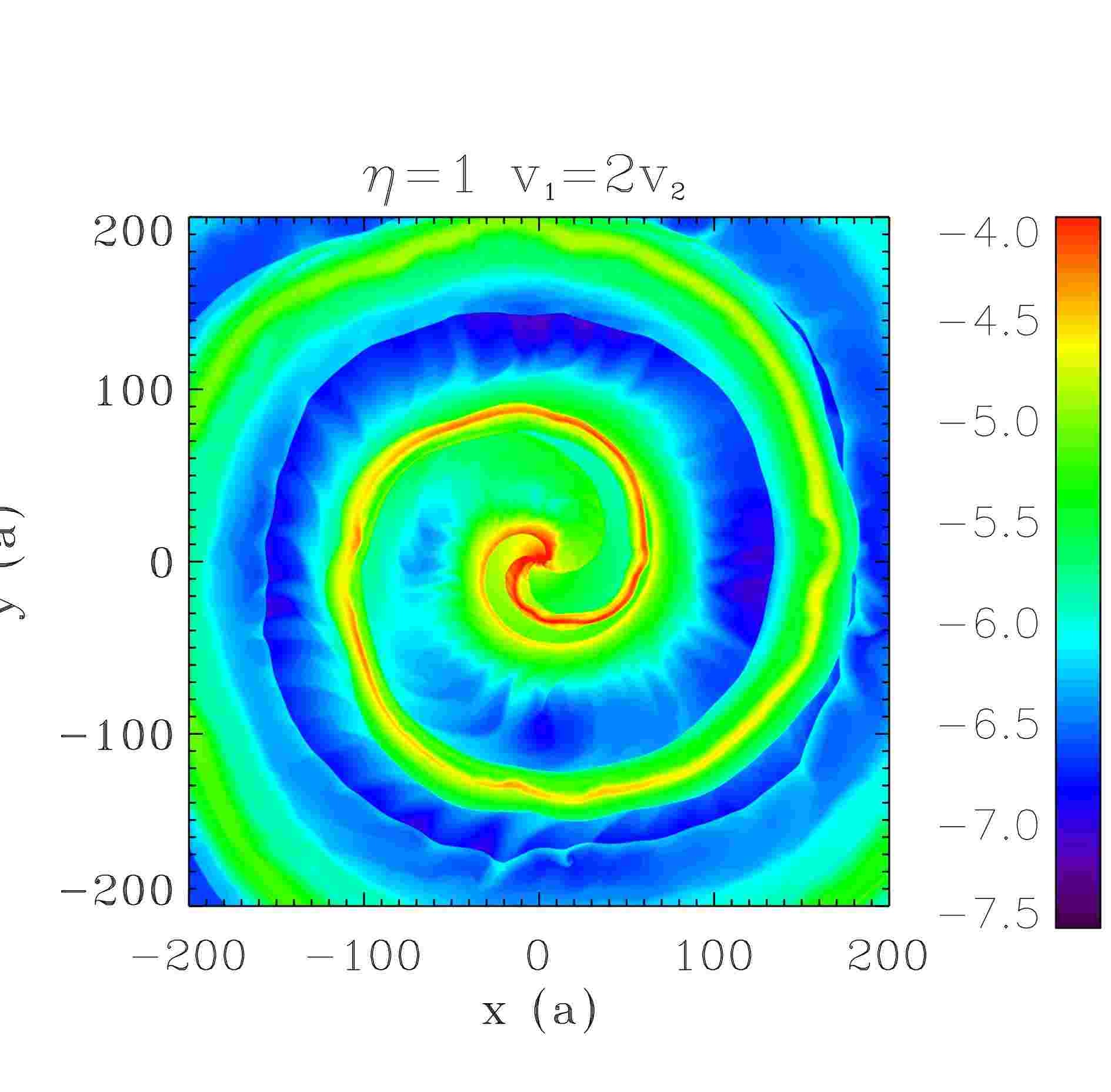}
  \includegraphics[width = .3\textwidth]{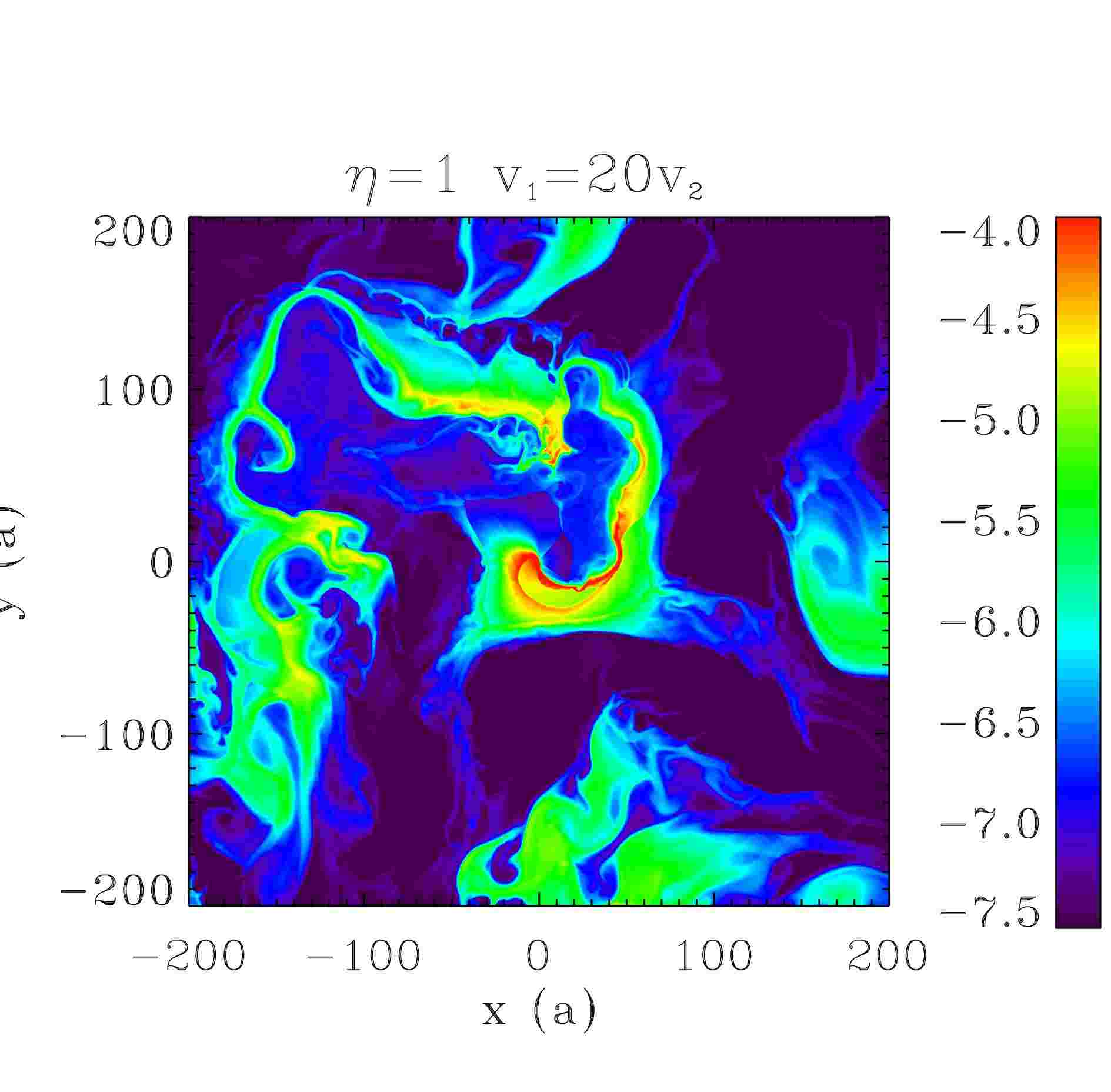}
  \includegraphics[width = .3\textwidth]{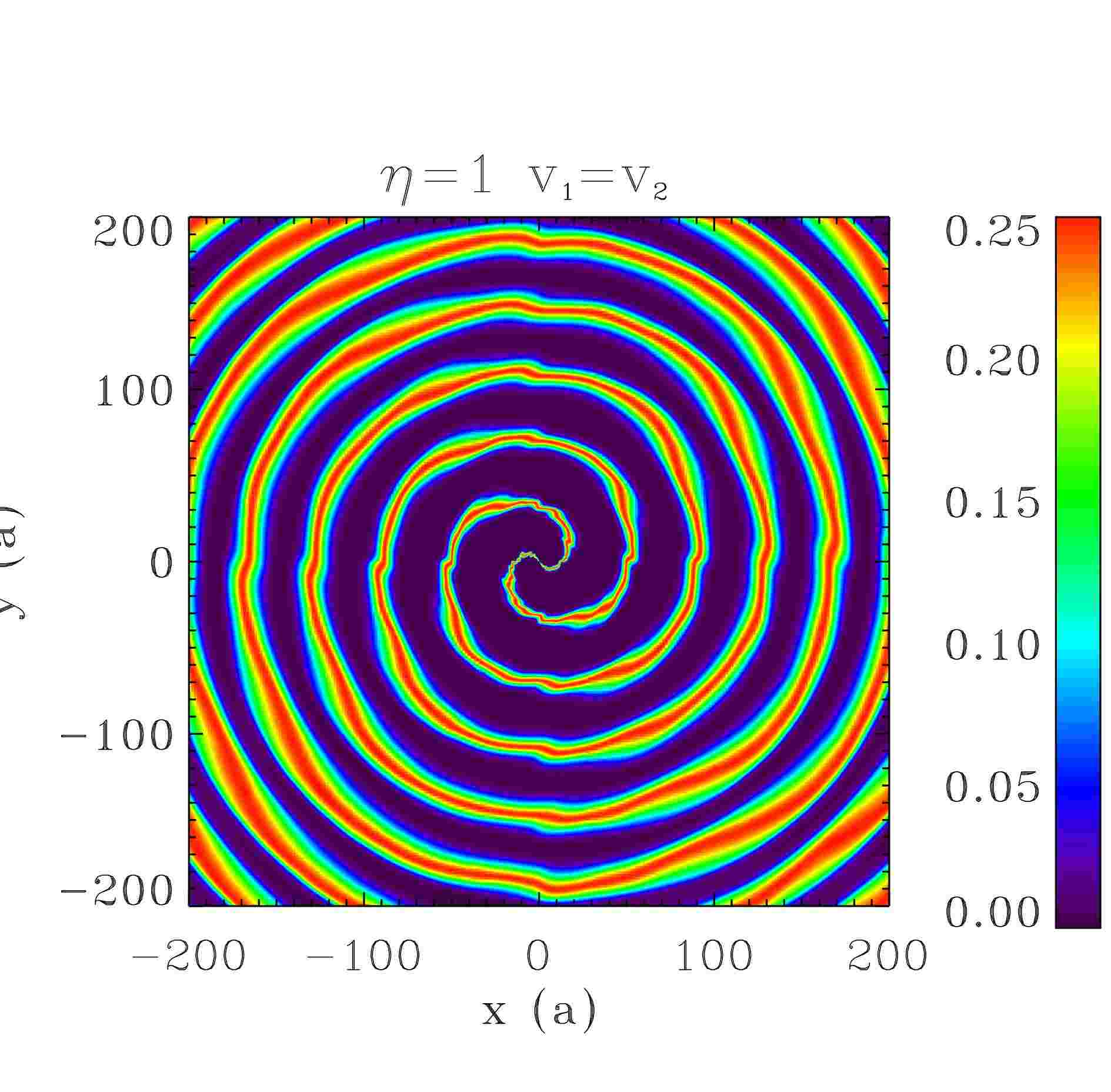}
  \includegraphics[width = .3\textwidth]{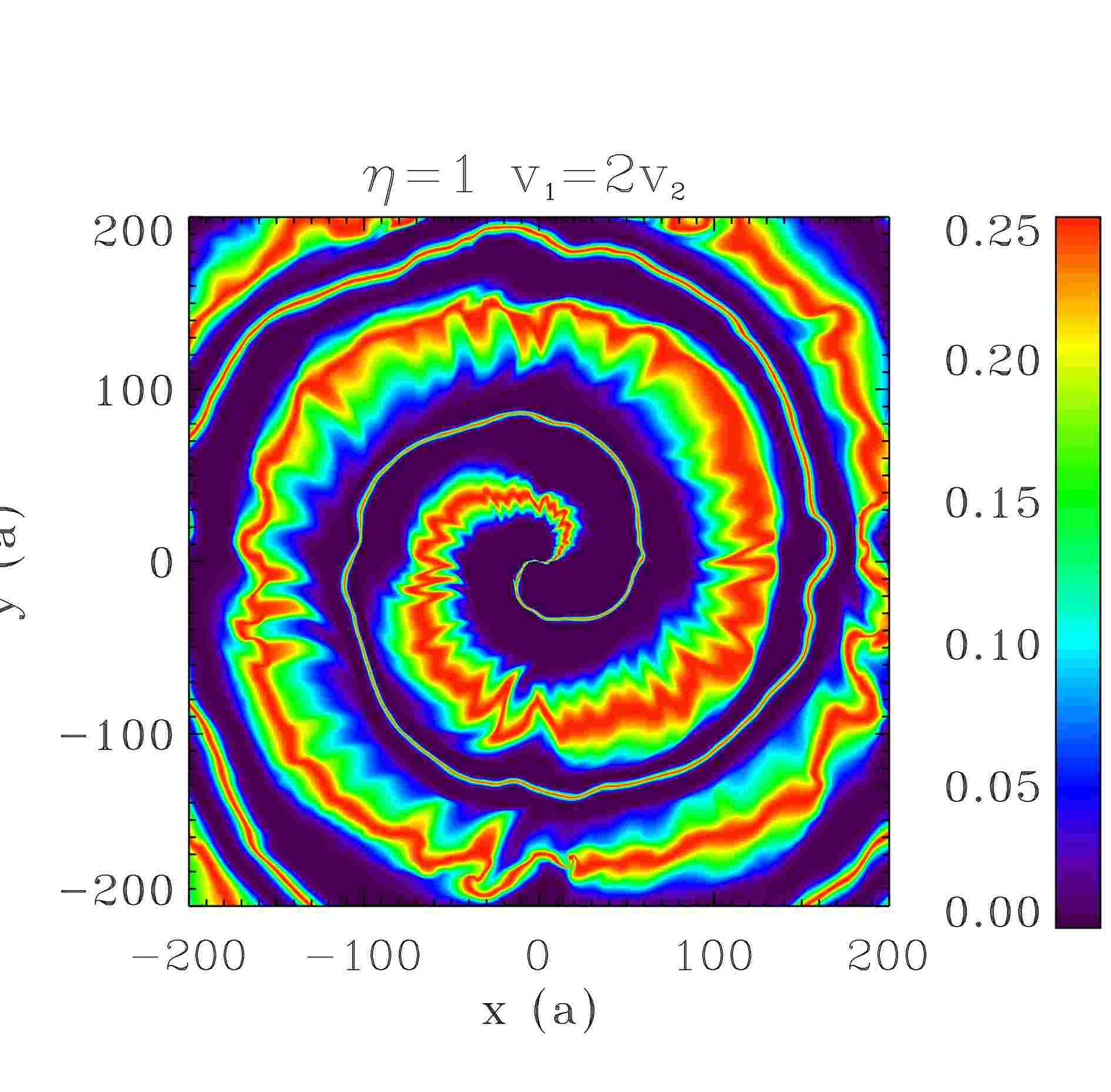}
  \includegraphics[width = .3\textwidth]{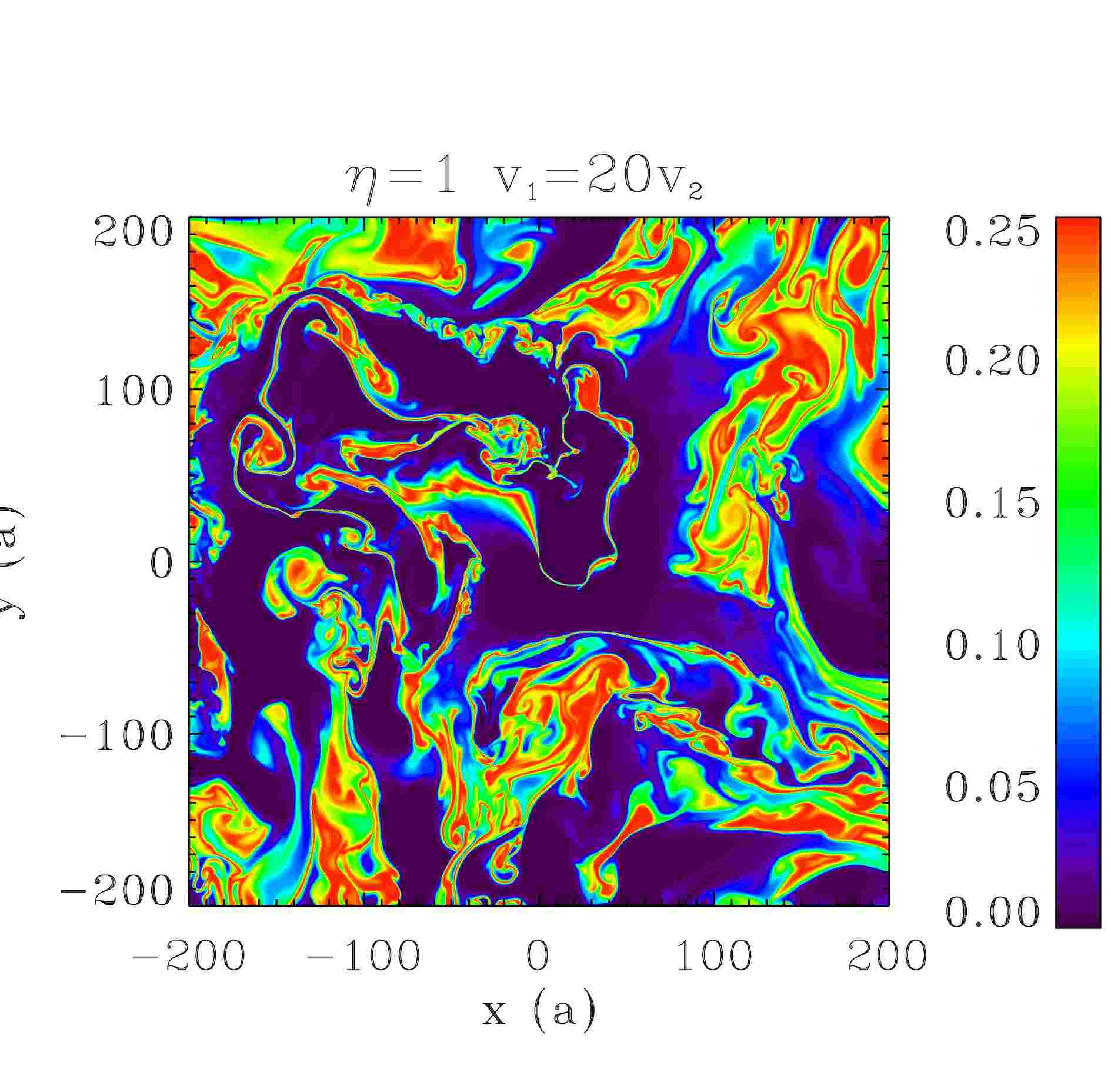}
  \caption{ Density (upper row)  and mixing (bottom row) for 2D simulations with increasing velocity ratios and decreasing density ratios (from left to right). On the left panel both winds are identical. The length scale is the binary separation. The maximum resolution is set by $l_{max}=15$.}
  \label{fig:fig1}
\end{figure}

Fig.\ref{fig:fig1} shows density and mixing for winds with equal moment fluxes ($\dot{M_1}v_1=\dot{M}_2v_2$) but increasing velocity ratios (and decreasing mass loss rate ratios). When both winds are identical (left panel), the spiral arms have the same step and same width. When the winds have dierent speeds, the spiral arm propagating into the faster, lower density wind expands, while the arm propagating into the denser and slower wind gets compressed. Mixing is more important in the wider arm \citep{PaperII}. When increasing the velocity dierence even more ($v_1=v_2 = 20$), the spiral structure is destroyed due to the KHI. Performing additional simulations, we show that an important density dierence can have a stabilising eect, even when winds have dierent speeds. We find that the step of the spiral is primarly set by $v\times  P$ where $ P$ is the orbital period of the system and $v$ the speed of the dominating wind, with significant corrections fromt he weaker wind in some cases.

\begin{figure}[h]
  \centering
  \includegraphics[width = .3\textwidth]{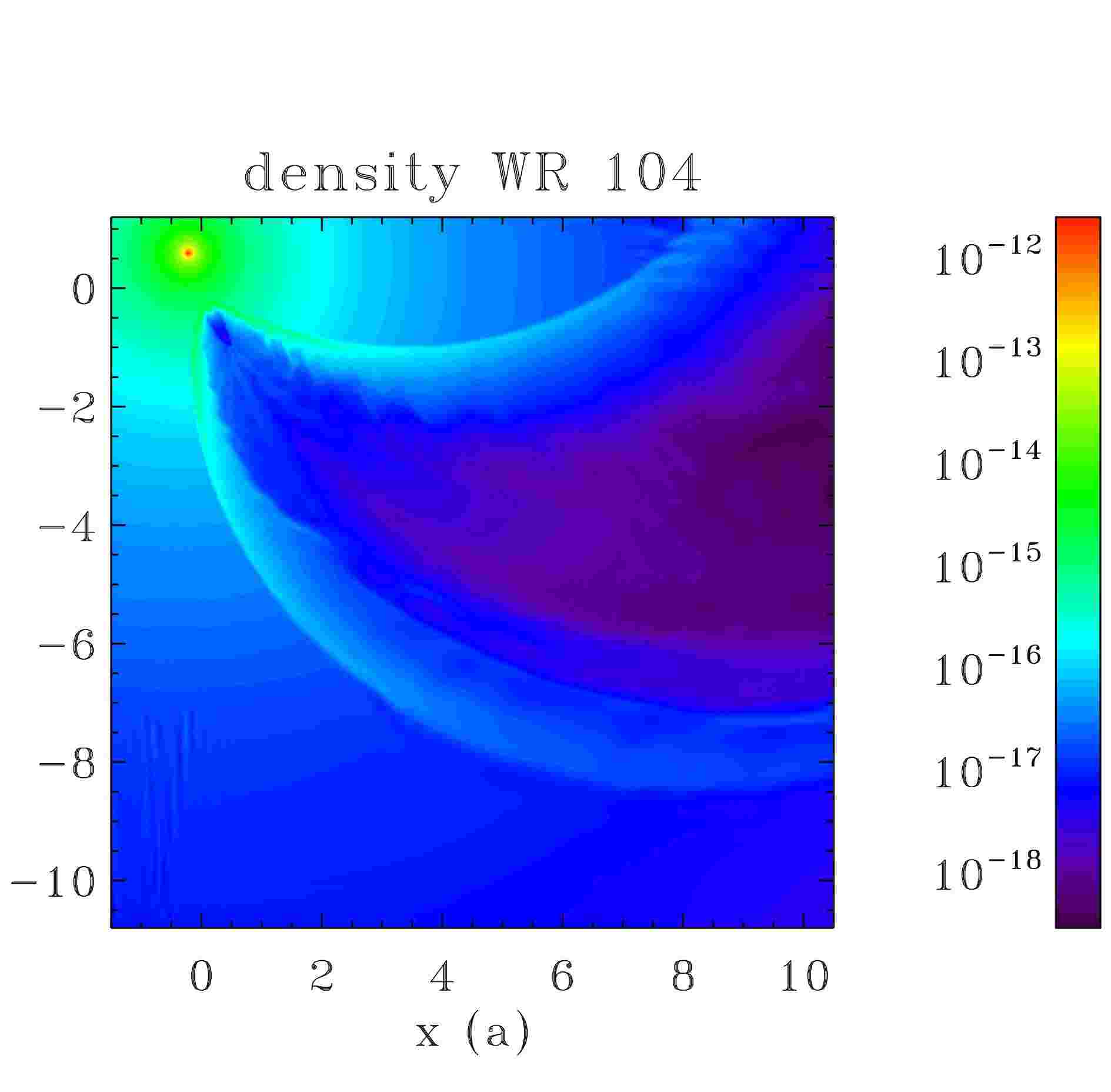}
  \includegraphics[width = .3\textwidth]{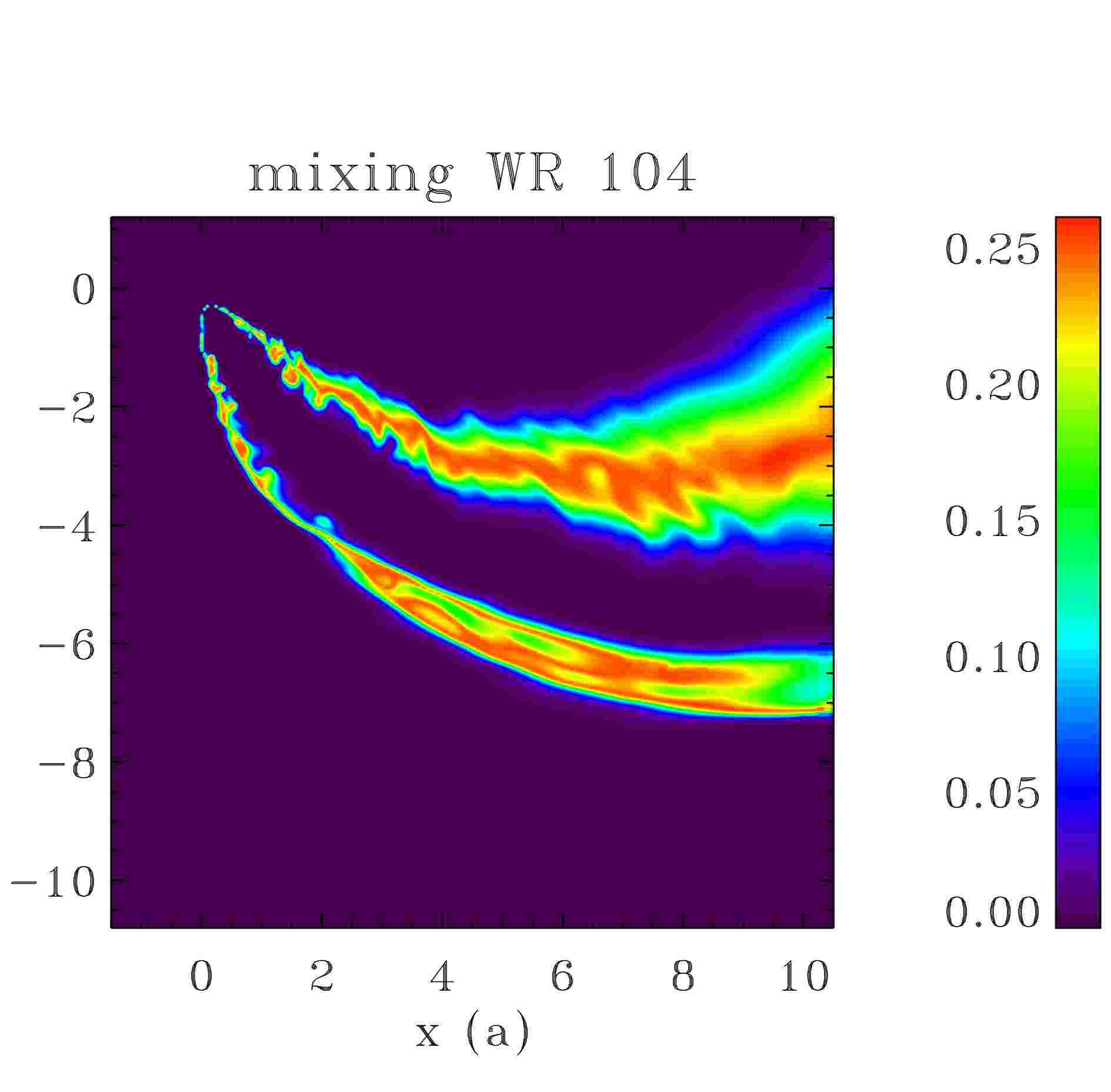}
  \includegraphics[width = .3\textwidth]{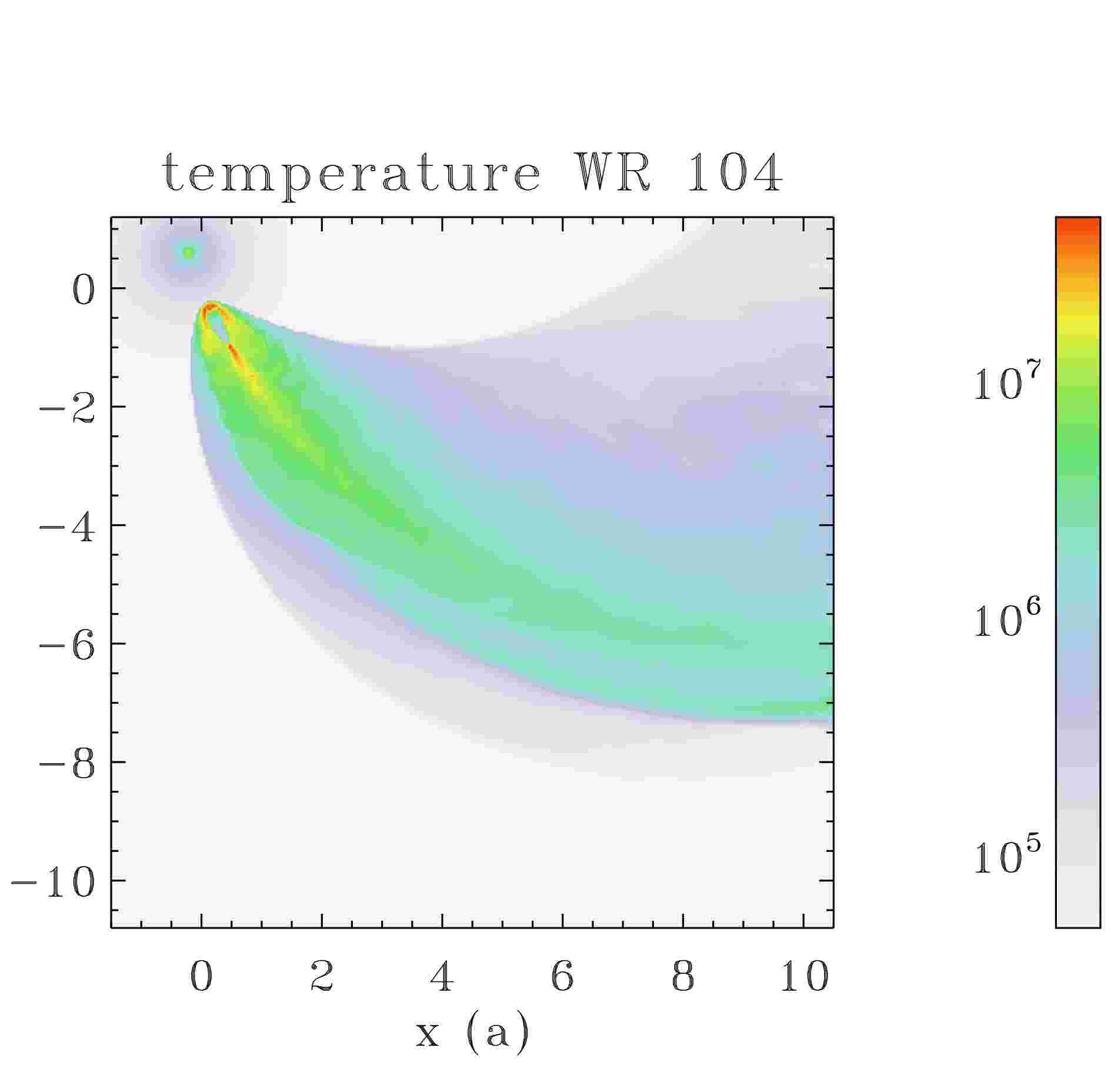}
  \caption{ Density (g cm$^{-3}$), mixing and temperature (K) in the orbital plane of the 3D simulation of WR 104. The length scale is the binary separation.}
  \label{fig:fig2}
\end{figure}

We perform a detailed study of WR 104. Our goal is to determine whether a hydrodynamic model with adiabatic winds can reproduce the observed spiral structure in WR 104. The WR wind is very hostile to dust formation because of its low density, high temperature and absence of hydrogen and the colliding wind region is likely to facilitate it \citep{2003IAUS..212..139W}. In WR 104 the wind from the WR strongly dominates the wind from the early-type star and the shocks form very close to the latter. A very high resolution is thus necessary close to the binary and is incompatible with a large scale 3D simulation. We thus perform a small scale 3D simulation and a complementary large 2D simulation. The 2D simulation confirms the Archimedean spiral and indicates the highest densities are reached at the edges of the spiral, suggesting they could be the location for dust formation (Lamberts et al. 2012). The results from the 3D simulation are given on Fig.\ref{fig:fig2}. The temperature map suggests temperature could become low enough to enable dust condensation ($\leq$ 6000 K) at half a turn of the spiral, which is compatible with observations. Mixing (middle panel) between the winds is likely to bring hydrogen and facilitate chemistry. Still, the density map shows that, even in the walls of the spiral, the density remains below the critical value for dust formation 
($\simeq 10^6$ cm$^{-3}$). An adiabatic model is thus unable to account for dust formation in WR 104. Cooling is probably present, and increases the compression ratio of the shocks. It might also trigger the non-linear thin shell instability, which results in small high density zones.

\section{Extension of RAMSES to relativistic hydrodynamics, application to $\gamma$-ray binaries}
The equations of RHD can be written as a system of conservation equations ($c\equiv 1$)\, :
\begin{equation}\label{eq:RHD}\nonumber
\begin{array}{ccc}
\frac{\partial{D}}{\partial{t}}+\frac{\partial{(Dv_j)}}{\partial{x_j}}&=&0\\
\frac{\partial{M_i}}{\partial{t}}+\frac{\partial{(M_iv_j+P\delta_{ij})}}{\partial{x_j}}&=&0\\
\frac{\partial{E}}{\partial{t}}+\frac{\partial{(E+P)v_j}}{\partial{x_j}}&=&0 
\end{array}
\quad
\textrm{with}
\quad
\left(
 \begin{array}{c} 
D \\ 
M_i\\
E 
\end{array}
\right)
=
\left(
\begin{array}{c}
\Gamma \rho \\ 
\Gamma^2 \rho h v_i\\ 
\Gamma^2\rho h -P
\end{array}
\right)
\end{equation}
where D is the mass density, $\mathbf{M}$ the momentum density and E the energy density in the frame of the laboratory.  The subscripts $i,j$ stand for the dimensions, $\delta_{i,j}$ is the Kronecker symbol. $h$ is the specific enthalpy, $\rho$ is the proper mass density, $v_i$ is the fluid three-velocity, $P$ is the gas pressure and $\gamma$ the adiabatic index. The Lorentz factor $\Gamma$ is given by 
\begin{equation}
\Gamma=\frac{1}{\sqrt{1-v^2}}
\end{equation}
These equations have a similar structure to the equations of hydrodynamics but are more complex to solve because strongly coupled to each other by the Lorentz factor and the enthalpy.  An additional numerical constraint arises from the the fact that the velocity must remain subluminal. The similarity with the equations of hydrodynamics  allows us to closely follow the algorithm implemented in RAMSES, performing  localised changes.

A first difficulty arises when passing from the conservative variables $(D, \mathcal{M}, E)^T$ to the primitive variables $(\rho,v,P)^T$ . There is no explicit solution, and a root finding algorithm is necessary to recover the primitive variables. Care has to be taken to avoid numerical problems in the ultrarelativistic and non-relativistic limits \citep{2007MNRAS.378.1118M}. Second-order precision is implemented in RAMSES following a MUSCL-Hancock method. The reconstruction of the primitive variables in space is not affected by special relativity but the prediction in time requires the determination of the Jacobian matrix of the system given by Eq. \ref{eq:RHD}. In case of a superluminal velocity in the reconstructed state, the code switches to a first order scheme. The relativistic summation of velocities changes the determination of the wavespeeds and the timestep.

In RAMSES, AMR is implemented following a tree-based method where parent cells are refined in child cells in a recursive structure \citep{1997ApJS..111...73K}. The child cells are gathered together in octs of size 2ndim. Interpolation from $l$ to $l + 1$ is done at the interface between dierent levels and when creating new refined cells. We perform the reconstruction on the conservative variables, and switch to first order when a non-physical state is found. A non-physical state occurs when the density or pressure are negative or the velocity superluminal. A physical state is guaranteed when ($D > 0, E^2 > D^2 + M^2$). Hydrodynamical updates are only performed at the highest level of refinement, and variables are computed at lower levels by averaging values over the child oct. Although each cell from the child oct at level $l$ satisfies ($D > 0, E^2 > D^2 + M^2$), it is not necessarily true on average, over the whole oct. This may lead to non-physical states at level $l-1$. Therefore, we choose to perform the restriction on the internal energy rather than the total energy. This method is currently implemented in RAMSES.

\begin{figure}[h]
  \centering
  \includegraphics[width = .3\textwidth]{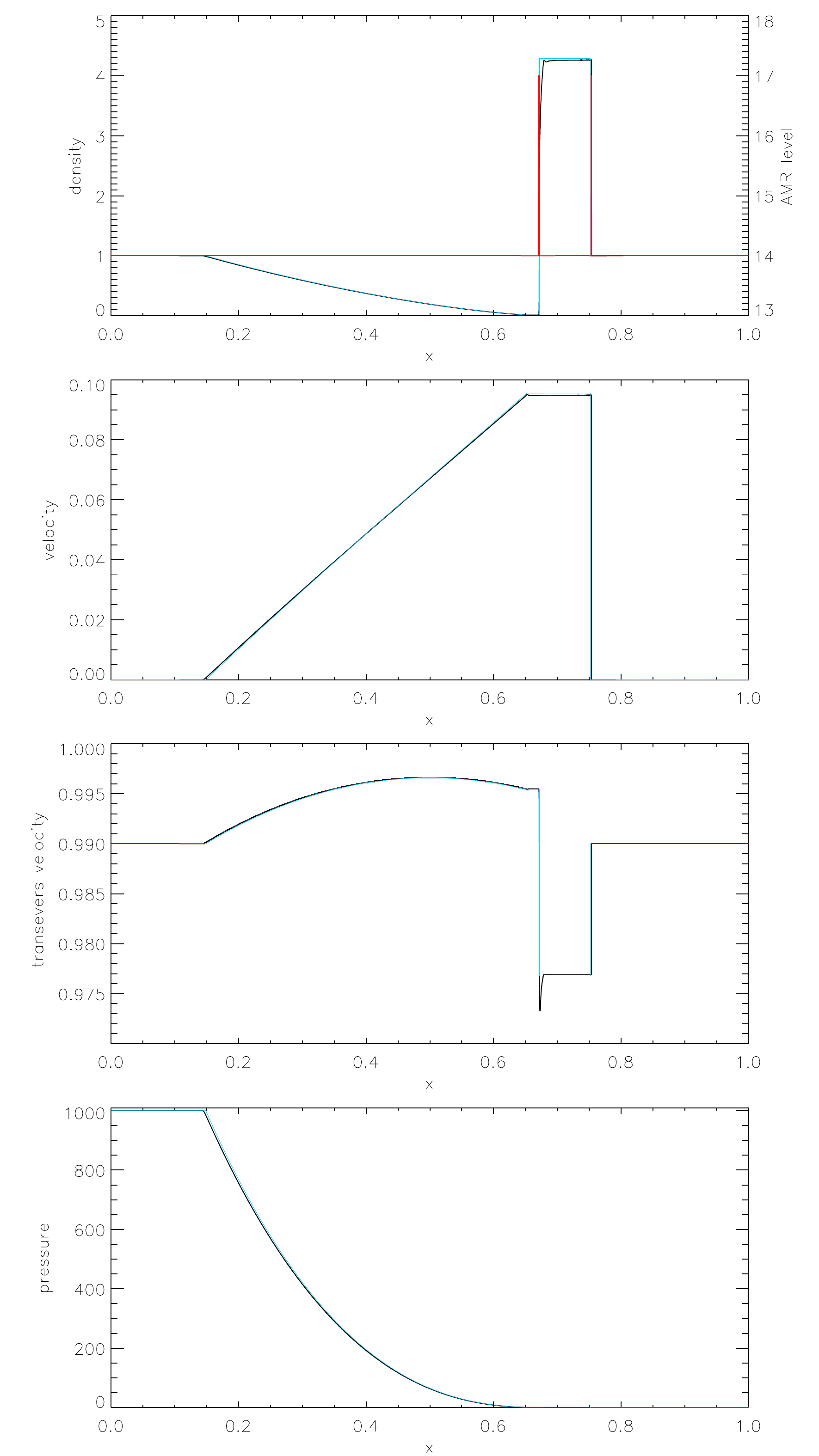}
  \includegraphics[width = .4\textwidth]{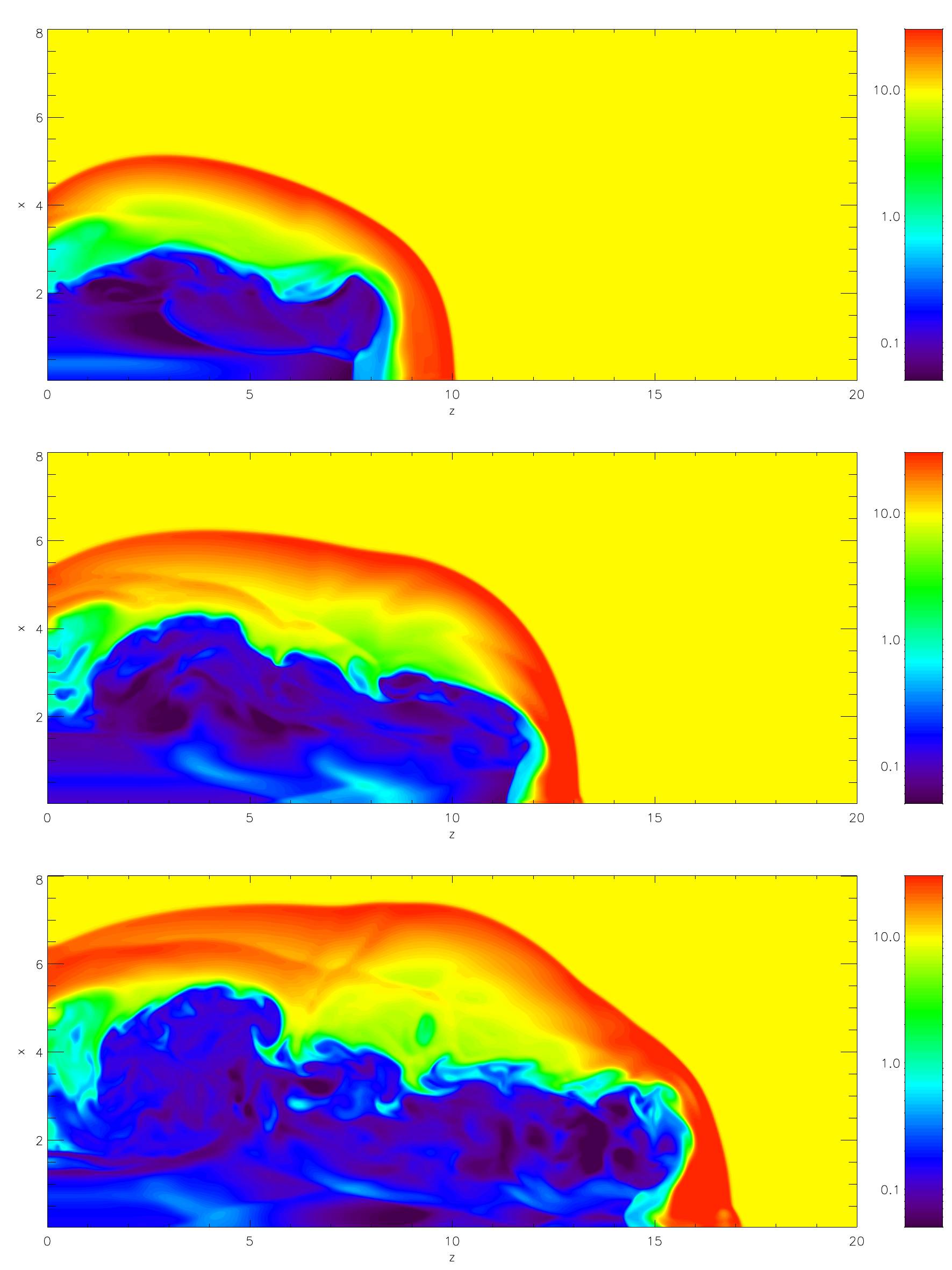}
  \caption{ Left panel : Sod test ($t=1.8$).Right panel : Simulation of the propagation of a 3D relativistic jet ($\Gamma_{max}=7.1$). From top to bottom : density at $t=20,30,40$ in a 3D jet starting from the left boundary of the domain}
  \label{fig:fig3}
\end{figure}

Fig. \ref{fig:fig3} shows two examples of numerical tests we performed with our new relativistic code. The left panel shows the density, transverse and parallel velocity and pressure in a one-dimensional Sod test. Contrary to the classical case, transverse velocities do impact the structure of shocks in RHD. The setup is taken from \citep{2006ApJS..166..410R} and constitutes a very stringent test due to the high Lorentz factor ($\Gamma_{max} = 120$) and important transverse velocity. A very high resolution (AMR levels shown in red) is needed to obtain a good agreement with the analytic solution (in blue). The right panel shows a the results of a 3D simulation of an axisymmetric jet following the setup by \citet{2002A&A...390.1177D}. Both tests show satisfactory results and indicate the code is ready for scientific use.

Our aim is to model $\gamma$-ray binaries. We focus on the small scale structure of the interaction between a stellar wind and a pulsar wind. The goal is to understand the impact of relativistic effects both on the structure and stability of the interaction region.We neglect orbital motion and focus on winds with equal moment fluxes. We perform preliminary simulations with various values of the momentum flux ratio and pulsar wind velocity. They prepare a large scale simulation to determine whether a stable structure is possible \citep{2012A&A...544A..59B}. Fig. 4 shows the density map for a simulation with the speed of the pulsar wind $v_p = 0.5$. The KHI develops in a similar fashion than in the classical case. The right panel shows the positions of the discontinuities for simulations with increasing values for the pulsar wind speed. The higher the value, the more the shocks are bent towards the star. This is a relativistic effect due to the impact of transverse velocities on the structure of shocks.

\begin{figure}[h]
  \centering
  \includegraphics[width = .3\textwidth]{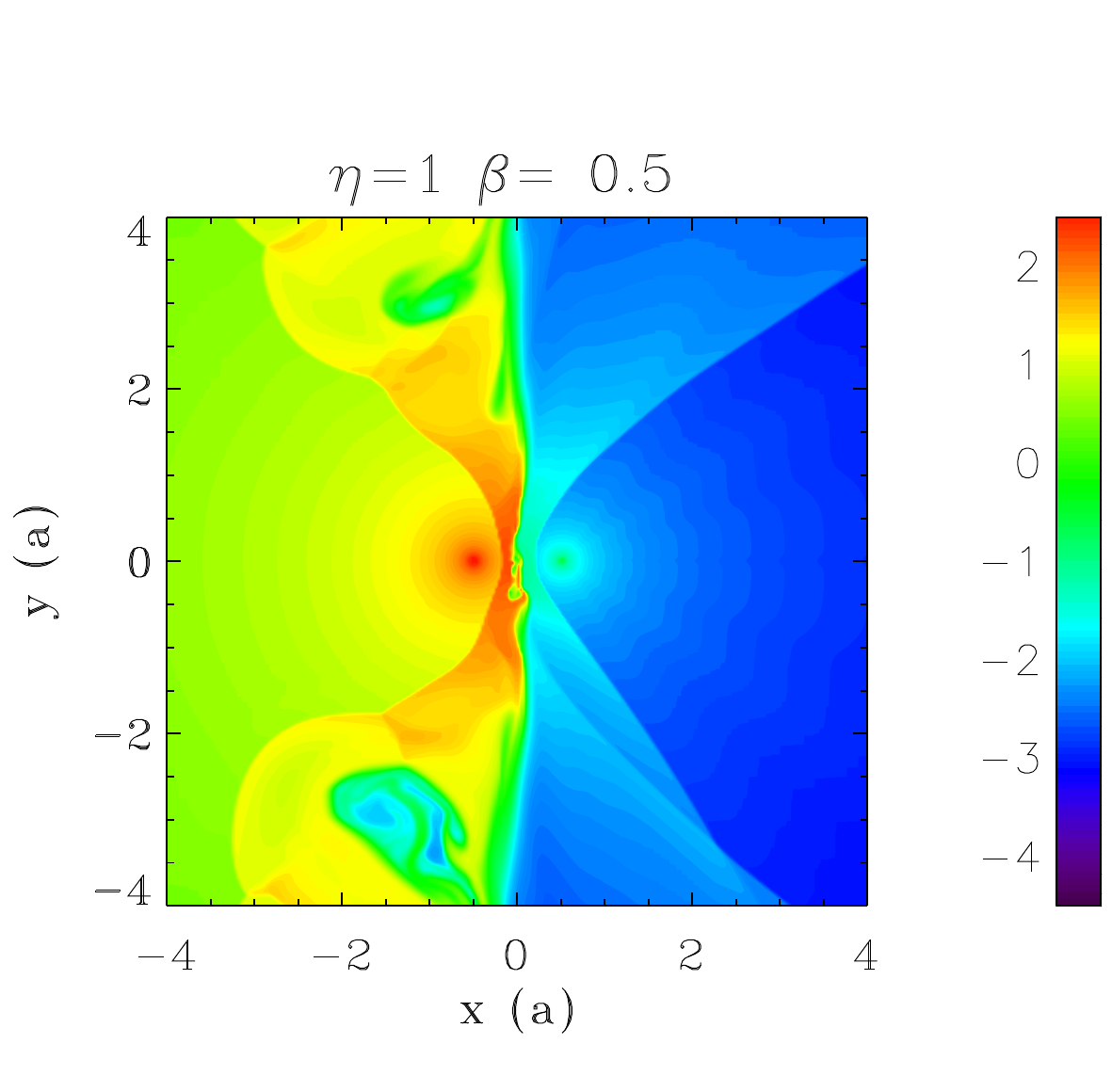}
  \includegraphics[width = .28\textwidth]{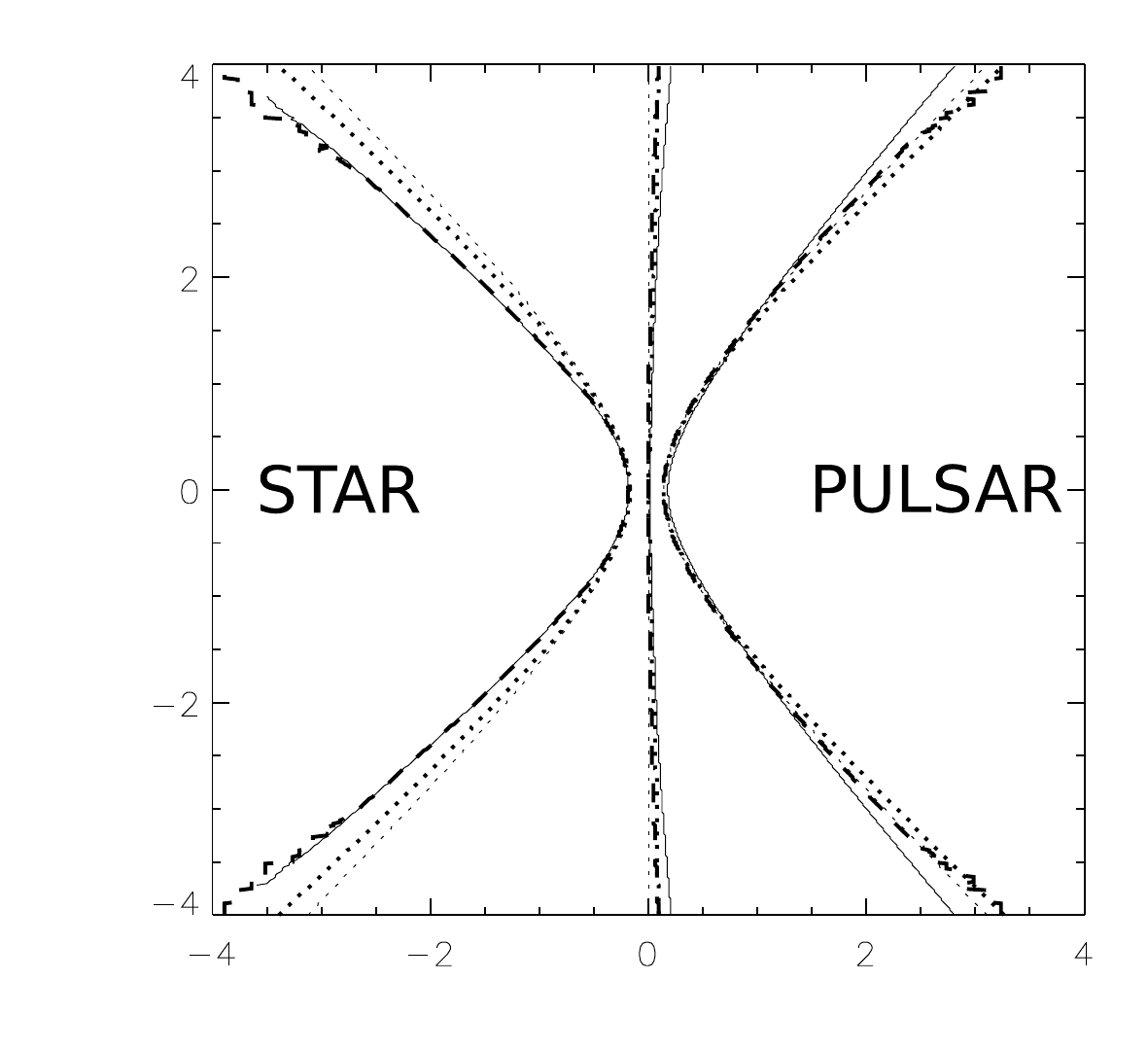}
  \caption{ Left panel : density map of a simulation with equal momentum flux, with$ v_p = 0,5$, the star is on the left, the pulsar on the right. Right panel : position of both shocks and the contact discontinuity, in simulations with different values for the velocity of the pulsar wind. We have $v_p= 0.01$ (thin dotted line), 0.1 (thick dotted lined), 0.5 (thick dashed line) and 0.9 (solid line).}
  \label{fig:fig4}
\end{figure}

\section{Conclusions and perspectives}
We performed high resolution simulations of colliding wind binaries at a spatial scale never reached before. We showed that the KHI may destroy the expected large scale structure. Simulations of WR 104 match well with the observed structure and indicate cooling has to be taken into account to allow dust formation in this system. To model $\gamma$-ray binaries, we extended RAMSES to relativistic hydrodynamics. Preliminary simulations of $\gamma$-ray binaries confirm a similar structure to stellar binaries. The relativistic extension of RAMSES allows the use of AMR and is suited for the study of gamma-ray bursts, relativistic jets or pulsar wind nebulae. It will be part of the next public release.

\begin{acknowledgments}
AL and GD are supported by the European Community via contract ERC-StG-200911. Calculations have been performed at CEA on the DAPHPC cluster and using HPC resources from GENCI- [CINES] (Grant 2011046391)
\end{acknowledgments}

\bibliography{biblio}

\end{document}